\documentstyle[aps,epsf]{revtex}                        

\draft
\preprint{}
\twocolumn

\title{Non--locality of particle spin:
a consequence of interaction energy?}

\author{W. A. Hofer} 
\address{Technische Universit\"at Wien\\
         A--1040 Vienna, Austria}

\begin{document}
\maketitle


\begin{abstract}
Neutron interference measurements with macroscopic beam separation
allow to study the influence of magnetic fields on spin properties.
By calculating the interaction energy with a dynamical
and deterministic model, we are able to establish that the phase
shift on one component of the neutron beam is linear with magnetic
intensity, and equally, that interaction energy as well as phase
shifts do not depend on the orientation of the magnetic field.
The theoretical treatment allows the conclusion that the non--local
properties of particle spin derive from the classical equation for
interaction energy $W = - \vec \mu \cdot \vec B$ and the fact, that
interaction energy does not depend on magnetic field orientation. 
Additionally, it can be established that the $4 \pi$ symmetry of
spinors in this case depends on the scaling of magnetic fields.
\end{abstract} 


Spin properties of particles, initially formalized for electron
states within the hydrogen atom by Goudsmit and Uhlenbeck 
\cite{GOU25}, are non--local and intrinsic features, since the
mathematical description by way of Pauli matrices does not allow
an identification of spin--orientation with defined directions
in space \cite{COT77}. As recently established, the direction of
spin can be related to the polarization of intrinsic magnetic fields
of particle propagation \cite{HOF96B}. The intrinsic magnetic fields
derive from the wave features of moving particles , which give rise 
to kinetic and electromagnetic potentials. The framework
suggested was shown to be an extension of classical electrodynamics
as well as quantum theory, since the treatment of micro physical
systems in either case is only a limited account of intrinsic particle
properties. The framework developed was applicable to electron as well
as photon propagation. A treatment of EPR--type measurements 
\cite{BEL64,ASP82} established,
additionally, that a significant correlation for a pair of spin
particles is likely to violate the uncertainty relations \cite{HOF96B}.

As the framework did not fully account for the non--local properties 
of spin, we may reconsider the problem in the context of actual 
measurement processes. More specifically the question, why measurements 
lead to the conclusion that spin {\em must be} a non--local property.
As the most conclusive experiments on spin properties currently are
neutron interference measurements with amplitude--splitting and
a beam separation in macroscopic dimensions, we apply the model 
developed to the intrinsic qualities of neutrons. The justification
for this extension of the original framework is to be seen in the
fact, that the results of measurements can be accounted for in
a purely classical framework of wave theory and x--ray interferences
\cite{RAU92}, which renders the fundamental equations of classical 
electrodynamics theoretically applicable. And that the relations of
classical electrodynamics, the Maxwell equations, are but a different
formulation of intrinsic particle properties, has already been 
proved \cite{HOF96B}.

In this paper we calculate, for the first time, the dynamical and
deterministic process of magnetic interactions, and it is shown
that the results are in accordance with interference measurements.
It will be established that non--locality of particle spin has its
origins in the qualities of the interaction process. The calculation
provides a reason, why interaction energy does not depend on the
orientation of magnetic fields. As a final result, we will show
that the $4 \pi$ symmetry of spinors, which was claimed to be proved
by these measurements, depends on the scaling of the magnetic field. 

\begin{figure}
\epsfxsize=1.0\hsize
\epsfbox{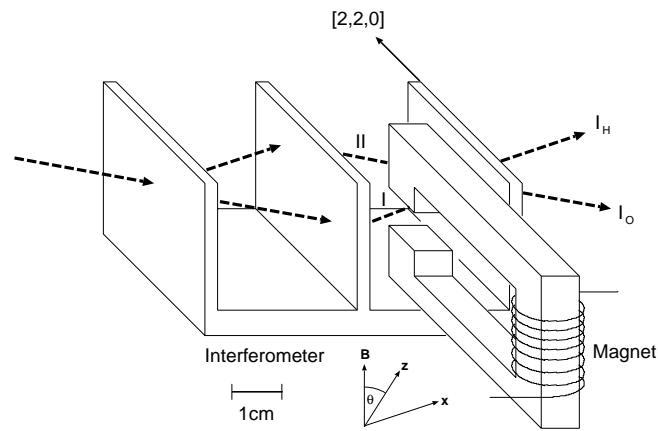}
\vspace{0.5cm}
\caption{Neutron interferometer for spin superposition measurement
according to Rauch. The incident neutron beam is split 
coherently at the first plate and reflected at the middle plate. The 
two beam components $I$ and $II$ are coherently superposed at the third 
plate of the interferometer, a static magnetic field in the path of 
beam $I$ leads to interference patterns depending on magnetic 
intensity}\label{sp001}
\end{figure}

The neutron interference experiments were performed in the Seventies
and Eighties by H. Rauch and A. Zeilinger at the Atomic Institute in 
Vienna \cite{RAU92,ZEI81,GREN84}. The experimental setup consisted of
a neutron source with a monochromator, the neutron beam of low 
amplitude was directed to an interferometer of perfect crystallic
properties. The intensity of the incident neutron beam was in every 
case such that only a single neutron passed the interferometer at a 
given moment. The first plane of the interferometer served as an amplitude
division device, the two separate beams (beam--separation in the range 
of cm) were reflected and finally recombind in the second and third plate. 
A static magnetic field in one path was used to alter the spin 
orientation of a single beam component, and the recombination of the
two separate beams then showed characteristic interference patterns
depending on the intensity of the magnetic field applied.
(see Fig. \ref{sp001}). 

For our theoretical model we postulate initially that neutrons  
possess wave like properties described by a wave function $\psi$ of 
single particles, intrinsic potentials to account for periodic mass 
distributions, and they shall be subject to the fundamental Planck and 
de Broglie relations \cite{HOF96B}.
Additionally we suppose that neutron mass in motion possesses an
intrinsic magnetic field of a specific orientation $\vartheta$, which
shall be perpendicular to the axis of particle motion $\vec u$. The
justification for these assumptions has to be seen, as already 
mentioned, in the theoretical description of interference measurements,
which are compatible with classical electrodynamics and thus the
intrinsic properties of particles.

The amplitude of the wave function and electromagnetic intensity are 
related by ($I_{qt}$ is the intensity due to quantum theory, $I_{em}$ 
the intensity obtained in classical electrodynamics) \cite{HOF96B}:

\begin{eqnarray}\label{sp01}
&I_{qt}& \propto \psi^{*} \psi \propto |\vec E|^2 \propto I_{em} 
\nonumber \\ &\Rightarrow& I_{qt} \propto I_{em}
\end{eqnarray}

The following calculation is a {\em local} and {\em deterministic} 
deduction of magnetic interactions, based on intrinsic electromagnetic 
fields and the field equations of electromagnetic properties as well as
intrinsic potentials \cite{HOF96B}:

\begin{eqnarray}\label{sp02}
\frac{1}{u^2} \frac{\partial \, \vec E}{\partial t} &=&
\nabla \times \vec B \qquad
- \frac{\partial \, \vec B}{\partial t} =
\nabla \times \vec E \nonumber \\
\phi_{em} &=& \frac{1}{2} 
\left( \frac{1}{u^2} \vec E^2 + \vec B^2\right)
\end{eqnarray}

We consider the change of intrinsic fields due to a constant external
magnetic field $\vec B_{ext}$, the field vectors shall be given by
(for convenience the orientation of intrinsic magnetic fields shall
denote the z--axis of our coordinate system):

\begin{eqnarray}\label{sp03}
\vec B &=& (0, 0, B_{0}) \cos(k_{0} x - \omega_{0} t) \nonumber \\
\vec E &=& (0, E_{0}, 0) \cos(k_{0} x - \omega_{0} t) \\
\vec B_{ext} &=& (0, - \sin \vartheta, \cos \vartheta) B_{ext}
\nonumber
\end{eqnarray}

Accounting for the dynamic qualities of the process by linear 
increase of the magnetic field $t \in [0,\tau]$, the internal
fields will be at $\tau$:

\begin{eqnarray}\label{sp04}
E_{y}' &=& E_{0} \cos(k_{0} x - \omega_{0} t) - 
B_{ext} \frac{\cos \vartheta}{\tau} x \nonumber \\
E_{z}' &=& - B_{ext} \frac{\sin \vartheta}{\tau} x \\
B_{y}' &=& - B_{ext} \sin \vartheta \nonumber \\
B_{z}' &=& B_{0} \cos(k_{0} x - \omega_{0} t)
+ B_{ext} \cos \vartheta \nonumber
\end{eqnarray}

Additional informations about the system can be inferred from the
relation between the variables $x$ and $t$ as well as from the
relation between amplitudes \cite{HOF96B}:

\begin{eqnarray}\label{sp05}
\frac{x}{\tau} = u_{0} \quad E_{0} = u_{0} B_{0}
\end{eqnarray}

The electromagnetic potential due to interaction with the
magnetic field is then given by:

\begin{eqnarray}\label{sp06}
2 \phi_{em} &=& \left[B_{0} \cos(k_{0}x - \omega_{0}t) - 
B_{ext} \cos \vartheta\right]^2 + \left[B_{ext}\sin \vartheta\right]^2
+ \nonumber \\
&+& \left[B_{0} \cos(k_{0}x - \omega_{0}t) + 
B_{ext} \cos \vartheta\right]^2 + \left[B_{ext}\sin \vartheta\right]^2
\nonumber 
\end{eqnarray}
\begin{eqnarray}\label{sp07}
\phi_{em} &=& B_{0}^2 \cos^2(k_{0}x - \omega_{0}t) + B_{ext}^2 
\end{eqnarray}

The result is interesting due to two features: 

\begin{itemize}
\item
The potential of interaction does not depend on the angle $\vartheta$
of the magnetic field.
\end{itemize}

It can therefore not be formalized as the scalar
product of an intrinsic magnetic moment $\vec \mu$ and an external
field $\vec B_{ext}$:

\begin{eqnarray}\label{sp08}
W \ne - \vec \mu \cdot \vec B_{ext} \qquad 
\vec \mu, \vec B_{ext} \in R^3
\end{eqnarray}

or only, if the magnetic moment is a non--local variable: the 
non--local definition of particle spin in quantum theory can therefore 
be seen as a different expression of an equivalent result. And the 
motivation for this definition has to be seen in the missing account 
of the deterministic and dynamic development of the intrinsic variables. 
The result confirms a conclusion already drawn by analyzing electron 
photon interactions \cite{HOF96B}: the framework of quantum theory is 
essentially limited to interactions, its logical implications only 
become obvious, if interaction processes are considered. 
In the context of particle spin it explains, why spin in quantum 
theory {\em cannot} be a local property: {\em because} interactions 
do not depend on the direction of field polarization.  

\begin{figure}
\epsfxsize=1.0\hsize
\epsfbox{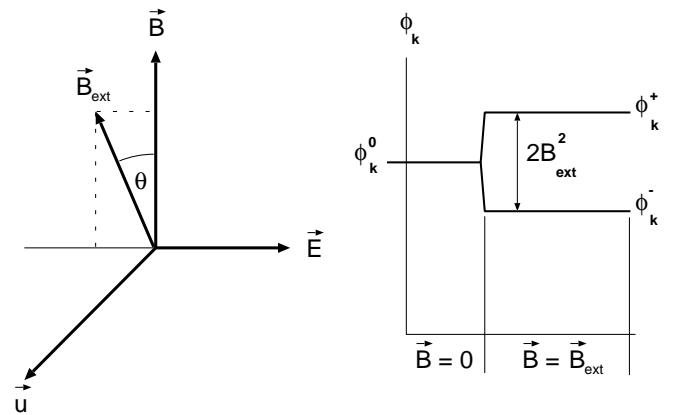}
\vspace{0.5cm}
\caption{Intrinsic electromagnetic fields and applied magnetostatic
field of the neutron beam (left). Shift of kinetic potential due
to interactions (right)}\label{sp002}
\end{figure}

\begin{itemize}
\item
The electromagnetic potential of the particle is higher than the 
original potential.
\end{itemize}

This result leaves two possibilities: either total energy density
of the particle remains constant -- which should be the case for 
neutral particles --, or the kinetic energy density of the 
particle is equally altered by interactions: which should apply
for charged particles. In both cases the kinetic potential during
magnetic interaction is changed, the alteration can be described by:

\begin{eqnarray}\label{sp09}
\phi_{k}' = \phi_{k}^0 \pm B_{ext}^2
\end{eqnarray} 

Intrinsic electromagnetic fields and kinetic potentials due to 
external magnetic fields are displayed in Fig. \ref{sp002}.
Due to the relations between the kinetic potential and density of
mass $\phi_{k} = \rho u^2$ and the relation between the wave function
and density of mass $\rho \propto \psi^2$ the properties of the
wave function in the region of interaction will equally be changed,
which means, that posterior superposition of the two separated beam
parts will yield a changed interference pattern. The easiest way to
calculate the changes in the affected beam is by estimating the 
difference of velocity. Since:

\begin{eqnarray} \label{sp10}
\langle \phi_{k}' - \phi_{k}^0 \rangle =: \triangle \phi_{k} =
- \bar \rho \,(\triangle u)^2 = - B_{ext}^2 
\end{eqnarray}

where $\bar \rho$ denotes denotes average density of the beam, as 
the wave length is much shorter than the macroscopic region of the
magnetic field in the interaction process, averaging is physically
justified. Then the phase difference $\alpha$ of the beam after 
$t_{1} = l/u_{0}$ seconds, where $l$ is the linear dimension of the
magnet, will be:

\begin{eqnarray}\label{sp11}
\alpha &=& 2 \pi \,\frac{\triangle u \cdot t_{1}}{\lambda} 
= 2 \pi \left(\frac{l}{\lambda}
\cdot \frac{B_{ext}}{\sqrt{\bar\rho \,u_{0}^2}} -  n \right) \nonumber \\ 
n &\in& N
\end{eqnarray}

\begin{figure}
\epsfxsize=1.0\hsize
\epsfbox{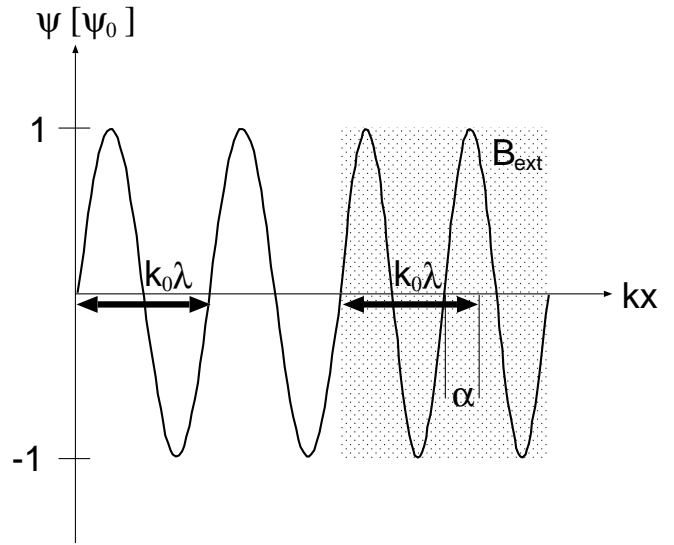}
\vspace{0.5cm}
\caption{Phase shift $\alpha$ in a magnetic field. The phase shift 
of the beam component $I$ depends on the scale of the external 
magnetic field $B_{ext}$}
\label{sp003}
\end{figure}

The theoretical result is consistent with the experimental result
by Rauch, that the phase of the beam is linear with the intensity
of the magnetic field applied \cite{RAU92}. That this phase shift
is sufficient for an experimental proof of the  $4 \pi$--symmetry 
of spinors seems to be a matter of convention, since it depends, 
essentially, on the scaling of the magnetic fields in terms
of kinetic potentials. All that can be inferred from measurements
is that magnetic fields affect the phase of the neutron beam, and
equally, that this effect does not depend on the orientation of the
magnetic field or the incident beam: both results are obtained in a
local and deterministic manner by this calculation. Fig. \ref{sp003} 
displays the changes of the wave function and the subsequent phase
shift due to magnetic interactions.

It should be noted, that the theoretical concept is only applicable
to monochromatic neutron beams. If particles have arbitrary energies 
then an equivalent theoretical framework also has to account for the 
phase shifts at different energy values: a close to classical 
interference pattern in this case cannot be expected.

Using the deterministic and causal model of intrinsic particle
properties we have, for the first time, calculated interactions
of particles in a magnetic field by evaluating the effect of
external and static magnetic fields on intrinsic particle
properties. The calculation was accomplished in a purely local
framework, and it was established that interaction energy does 
not depend on the orientation of the magnetic field. The classical
description of interaction energy by way of magnetic moments was 
shown to be unsuitable to account for the results achieved, and
it was found that the $4 \pi$ symmetry of spinors depends on the 
scaling of magnetic intensities.

The consequences of this new result for the framework of quantum theory
seem substantial: as the theoretical calculation established, the
quality of spin appears to be a simplification which does not
allow for a precise treatment of interaction processes. As previous
results, furthermore, revealed that the concept in itself is 
theoretically questionable \cite{HOF96B}, the result suggests a
theoretical revision of magnetic interactions in micro physics
based on deterministic and dynamical models.



\end{document}